# A Flexible and Secure Deployment Framework for Distributed Applications

Alan Dearle, Graham Kirby, Andrew McCarthy, and Juan Carlos Diaz y Carballo

School of Computer Science, University of St Andrews,
North Haugh, St Andrews, Fife KY16 9SS, Scotland, U.K.
{al, graham, ajm, jcd}@dcs.st-and.ac.uk

**Abstract.** This paper describes an implemented system that is designed to support the deployment of applications offering distributed services, comprising a number of distributed components. This is achieved by creating high level placement and topology descriptions that drive tools to deploy applications consisting of components running on multiple hosts. The system addresses issues of heterogeneity by providing abstractions over host-specific attributes yielding a homogeneous run-time environment into which components may be deployed. The run-time environments provide secure binding mechanisms that permit deployed components to bind to stored data and services on the hosts on which they are running.

## 1 Introduction

This paper describes an implemented system that is designed to support the deployment of applications offering distributed services, comprising a number of distributed components. A number of requirements for flexible service deployment may be identified, including:

- an architectural description of software components, the hosts on which they are to execute, and the interconnections between them [1]
- the ability to enact the architectural description to obtain a running deployment consisting of the specified set of components—requiring:
  - the ability to install and execute code on remote hosts
  - a security mechanism to prevent malicious parties from deploying and executing harmful agents, and deployed components from interfering with each other, either accidentally or maliciously
- support for component implementation using standard programming languages and appropriate programming models
- the ability for components to interface with off-the-shelf (COTS) components already deployed

Clearly, security considerations are a major issue in any flexible deployment infrastructure. Our system introduces new security domains, called *thin servers*, which can be placed within an existing network. Thin servers permit flexible and dynamic placement of both code and data by authorised users, in a secure and simple manner. They support a model of global computation in which objects have global identity, and the programmer may define the physical domain in which code is executed. Thin servers do not replace existing hosts, but are instead used to complement existing infrastructure to increase its usability and effectiveness. Indeed, thin servers can be co-hosted on conventional servers, and the services offered by them may be indistinguishable from conventional services.

In order to permit deployed components to be assembled into appropriate topologies and communicate with each other, the components must exhibit some degree of interface standardisation. In the implementation described here, communication is via asynchronous channels that may be dynamically rebound to arbitrary components either by the components themselves or by suitably privileged external parties.

All software applications are subject to evolutionary pressure. In order to respond to these pressures, it may be necessary to adapt parameters of the deployed services, including, but not limited to, the placement of components and data on machines and the components' interconnection topology.

This paper describes a framework that permits distributed services to be described, deployed and evolved in distributed contexts. It provides binding mechanisms that permit components to bind to local code, data and processes, including inter-node, inter-component bindings. When combined, these provide a run-time environment within which a deployed application may evolve. The framework contributes to the state of the art in six areas, providing:

1. mechanisms for deploying code and data in a distributed environment
2. abstractions over node specific attributes yielding a homogeneous run-time environment for deployed components
3. safe binding mechanisms so that deployed components can bind to stored data and services on the nodes on which they are running
4. mechanisms for describing and deploying distributed applications consisting of components running on multiple nodes
5. the ability to evolve the topology of deployed applications and components
6. security mechanisms that permit a wide range of policies to be implemented ranging from liberal to draconian

The deployment framework described here is based on an enabling infrastructure called Cingal[1] [2, 3]. Cingal itself addresses points 1-3 above, while the deployment infrastructure adds support for points 4-6.

## 2 The Cingal Computational Model

Cingal supports a conceptually simple computational model in which each thin server provides the following:
- a port to which a *bundle* of code and data may be sent for execution
- authentication mechanisms, preventing unauthorised code from executing

---

[1] *Computation IN Geographically Appropriate Locations*

- a content addressable store
- symbolic name binders for data and processes
- an extensible collection of execution environments called *machines*
- channel-based asynchronous inter-machine communication
- a capability system controlling access to stored data, machines and bindings.

The computational model is illustrated in **Fig 1**, which shows two hosts: a conventional host that might be running Windows or MacOS, and a thin server running the Cingal infrastructure. In order to execute code on a thin server, an OS process on the conventional host sends a bundle of code and data to the thin server where it is received by a daemon known as the *fire daemon*.

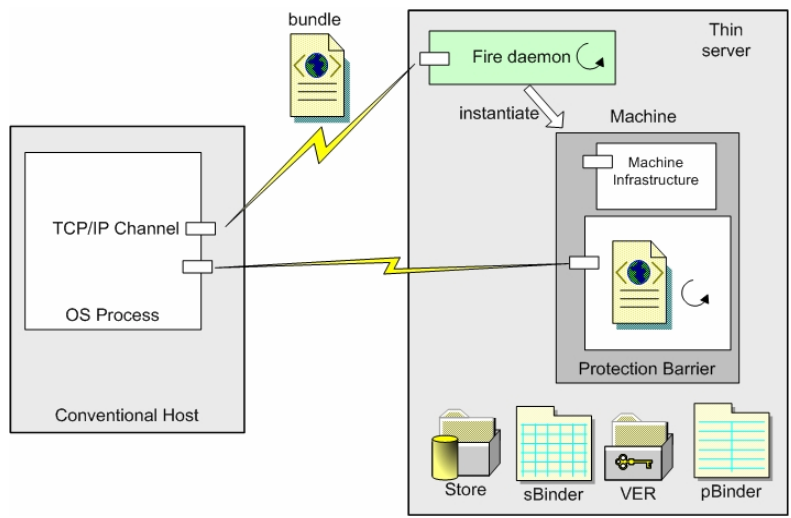

**Fig 1:** Cingal computational model

The fire daemon authenticates the bundle using mechanisms described later and, provided that it is authenticated, the bundle is *fired*. This causes a new operating system process to be created, which executes the code in the bundle. This process, which we term a *machine*, contains the code in the bundle and *machine infrastructure* containing code and data structures provided by the thin server. The infrastructure provides mechanisms to allow executing bundles to access the services provided by thin servers, to interface with the protection mechanisms, and to permit inter-machine communication channels to be established. The infrastructure may be invoked from other processes via an interface called the *machine channel* and from the executing bundle via an API (called the *machine API*) provided to the bundle when it is initialised[2].

The thin server infrastructure includes a number of services that may be invoked from bundles executing within machines. These are the *store*, *store binder*, *process binder* and *valid entity repository* (VER), providing storage, binding and certificate storage respectively. A running bundle's interactions with these services are restricted via a capability protection scheme mediated via library code in the infrastructure.

---

[2] API documentation is available at *http://www-systems.dcs.st-and.ac.uk/cingal/docs/*.

The *bundle* is the only user-level entity that may be executed in Cingal. It is passive, consisting of a closure of code and data and a set of bindings naming the data. In the current implementation bundles are XML-encoded, as illustrated in **Fig 2**. Each bundle carries an *authentication* element with attributes *entity* and *signature*. The *entity* identifies the bundle using a globally unique identifier (GUID) implemented via an MD5 key. The *signature* is used by the security infrastructure. In the current implementation, code may be either MIME-encoded Java classes or JavaScript source. In principle, any programming language could be used for encoding components, provided that the appropriate run-time support was provided. When a bundle is fired, execution begins at the entry point specified by the *entry* attribute of the code element, which specifies code that implements a standard interface. The data section of a bundle, known as its *payload*, comprises data with each datum having a unique *id* attribute. In the example the bundle carries one datum named *ToDoList*. It is common for bundles to carry other bundles in their payload, in order to install bundles in the store or fire them in other machines.

Subject to the capability protection scheme, bundles may carry out any arbitrary computation that they are encoded to perform, including the provision of network services.

```xml
<BUNDLE>
    <AUTHENTICATION    entity="19730129df7447eb91509"
                       signature="DQoew3rasZ...9wu9ySLGU"/>
    <CODE entry="uk.ac.stand.cingal.Runner" type="java">
        <CLASS name="uk.ac.stand.cingal.Runner">
            MamF2YS9sYW5nL09ia…
        </CLASS>
    </CODE>
    <DATA><DATUM id="ToDoList">
         <TODOLIST>
            <TASK guid="urn:cingal:325444" type="RUN">
               <DATUM id="StoreGuid">
                   Lvcxk3wnAIUN…
               </DATUM>
            </TASK>
         </TODOLIST>
     </DATUM></DATA>
</BUNDLE>
```

**Fig 2**: An example bundle

The store provided by each machine is a collection of passive data and supports the storage of arbitrary bundles. So that a bundle may be retrieved, a *key* in the form of a GUID is returned by the store on its insertion. If that key is later presented, the original bundle is returned. Stores do not support any update operations. Where the effects of update are required by an application, these may be obtained using *binders*.

Cingal thin servers provide two kinds of symbolic name binders: a *store binder* for naming entities in the store and a *process binder* for naming active machines. Both provide the ability to manipulate bindings though symbolic names and provide standard *put*, *remove* and *get* operations.

Cingal supports asynchronous message-oriented inter-machine communication. All communication is via *channels* that support conventional *read* and *write* operations. Each machine has associated with it a minimum of two channels as shown in **Fig 1**. The first is called the *machine channel* and is used to communicate with the machine infrastructure. The second is called the *default channel* and is used to communicate with the bundle running within the machine. An interface to the default channel is returned to its progenitor whenever a bundle is fired. The fired bundle may access the default channel via the machine API.

The channel established between a bundle and its progenitor is normally used for diagnostics and the passing of parameters. In order to accommodate change and dynamic deployment, the Cingal computational model supports named channels between entities. This idea stems from Milner's π-calculus [4]. Using named channels, individual executing bundles are isolated from the specifics of what components are connected to them. This isolation permits channels to be connected, disconnected and reconnected independently of the running program. Connections may be manipulated by the connected bundles or by third parties. This ability is necessary for the orchestration, evolution and autonomic management of deployed applications [5].

Within the machine infrastructure a component called the *connection manager* is responsible for the management of named channels. It maintains an associative mapping of names to channels. This mapping may be manipulated by other machines via the machine's machine port and by the bundle being executed via the machine API.

The model described thus far is a perfect virus propagation mechanism. Code may be executed on remote nodes and that code may create new processes, update the store, create name bindings and fire bundles on other thin servers. Cingal implements a two-level protection system. The first level of security restriction is on the firing of bundles. A conventional Unix or Windows style security model is not appropriate for thin servers, which do not have users in the conventional sense. Instead, security is achieved by means of digital signatures and certificates. Each thin server maintains a list of trusted *entities*, each associated with a security certificate. Entities might correspond to organisations, humans or other thin servers. This data structure is maintained by a process called the *Valid Entity Repository* (VER).

Bundles presented for firing from outwith a thin server must be signed by a valid entity stored in the VER. The VER maintains an associative data structure indexed by the entity *id* and mapping to a tuple including certificates and rights. Operations are provided for adding and removing entities from the repository. Of course these operations are subject to the second protection mechanism, which is capability-based. An example of a signed bundle was shown in **Fig 2**.

The *entity* attribute of the *authentication* element represents the name of an entity in the VER of the thin server on which the bundle is being fired and the *signature* is the signed body of the bundle's code payload. The thin server deployment infrastructure for deploying bundles from conventional machines provides programmers with methods that simplify the signing of bundles.

The signing of bundles and their authentication on arrival at thin servers prevents the misuse of thin servers by unauthorised entities. However it does not prevent a bundle from interfering with other bundles or entities in the binder or store. It is possible for bundles to be totally isolated, giving the illusion that each is the only entity running on a thin server. Conversely, bundles may share resources when appropriate.

To address these needs, the second protection mechanism provided by thin servers is capability-based. In addition to the signatures stored in the VER, thin servers store segregated capabilities for entities stored in the *store*, *sBinder*, *pBinder* and the VER itself. Whenever a running bundle attempts an operation, the capabilities stored in the VER associated with the entity that invoked the operation are checked. The operation only proceeds if the entity holds sufficient privilege.

## 3 Application Deployment

The Cingal system provides the infrastructure for deploying components on arbitrary suitably enabled hosts. However, additional infrastructure is needed to *a)* describe distributed architectures, and *b)* deploy components from the descriptions. This infrastructure comprises a *description language*, a *deployment engine*, and various mobile code documents and tools.

The description language is an XML schema, instances of which are *Deployment Description Documents* (DDDs). Each DDD contains an architectural description of an application, comprising a set of autonomous software components, the hosts on which they are to execute, and the interconnections between them.

The deployment engine takes a DDD as input and deploys the components described in it on the appropriate hosts. These components are pushed to the hosts as Cingal bundles; every participating host is Cingal-enabled. The deployment engine also pushes various tools to the hosts to carry out local deployment tasks *in situ*, principally installing and initialising the components, and configuring the interconnection topology of the deployed application. These tools are also transferred as bundles.

The data element of a tool bundle includes a control document called a *to-do list*. This contains a set of tasks to be attempted by the tool when it arrives and is fired on the destination thin server. When the tool completes these tasks, it sends a *task report* document back to the deployment engine, listing the outcomes of each task and any other associated information. Examples of associated information might include the GUIDs of stored bundles, or the names of channels and machines.

The three primary tools are *installers*, *runners* and *wirers*. An *installer* installs an arbitrary number of payload bundles into the store of the destination thin server. A *runner* starts the execution of a number of bundles previously installed in the store. A *wirer* is responsible for making concrete connections between pairs of components using the named channel mechanism.

Under control of these tools, each application component on a thin server moves between the following states:

- **installed**: when the bundle has been installed into the store
- **running**: when the bundle has been fired and started computation; any reads or writes on named channels will block since they are not connected
- **wired**: when the bundle has started computation and all named channels have been connected to other components

During initial deployment of an application the constituent components move from *installed* to *running* to *wired*. During subsequent evolution the components may move from *wired* to *running* to *wired* again, in cases where only the interconnection topology needs to change, or from *wired* to *running* to *installed* to *running* to *wired* again, in cases where components need to be moved to different hosts in the network.

Each instance of a deployment tool is pushed to the appropriate host as a bundle containing a fixed code element depending on its type (i.e. installer, runner or wirer), and a data element configured to its particular role. Thus every installer bundle contains the same generic installer implementation (currently a Java class), which is specialised by the bundle payload—and similarly for runners and wirers.

The example installer bundle shown in **Fig 3** contains the installer code: the class *uk.ac.stand.cingal.Installer*. The payload carries another bundle, itself containing the classes *Server* and *CacheUpdater*, and a *to-do list* specifying that that bundle (identified by the *id* attribute value `"urn:cingal:a222jdjd2s"`) should be installed.

```xml
<BUNDLE>
   <AUTHENTICATION    entity="197301m7wWwrPxX9..EySLGU"
                      signature="kUdzrv6T..fFNn5Kap" />
   <CODE entry="uk.ac.stand.cingal.Installer" type="java">
      <CLASS name="uk.ac.stand.cingal.Installer">
         5leLKJJbnQBAAMoKU...
      </CLASS>
   </CODE>
   <DATA>
      <DATUM id="urn:cingal:a222jdjd2s">
         <BUNDLE>
            <AUTHENTICATION    entity="1973012..91509"
                               signature="DQowLAIUNs..if1Dn5Kap" />
            <CODE entry="Server" type="java">
               <CLASS name="Server">
                  5lHRHAJMnQDD43MoKU...
               </CLASS>
               <CLASS name="CacheUpdater">
                  5leHdkvjidfjFFFDDEEU...
               </CLASS>
            </CODE>
            <DATA />
         </BUNDLE>
      </DATUM>
      <DATUM id="ToDoList">
         <TODOLIST>
            <TASK guid="urn:cingal:aEcncdeEe" type="INSTALL">
               <DATUM id="PayloadRef">
                  urn:cingal:a222jdjd2s
               </DATUM>
            </TASK>
         </TODOLIST>
      </DATUM>
   </DATA>
</BUNDLE>
```

**Fig 3**: An installer bundle

Examples of runners and wirers are shown in **Figs 2** and **6** respectively.

A DDD is a static description of a distributed graph of components; an example is shown in **Fig 4**. It specifies the locations of the required components (bundles), the hosts available, the mapping of components to hosts (deployments) and the connections between named channel pairs.

```xml
<DDD name="ServerAndCacheApplication">
   <BUNDLES>
      <BUNDLE name="Server" source="file://C:\bundles\server.xml" />
      <BUNDLE name="Cache" source="file://C:\bundles\cache.xml" />
   </BUNDLES>
   <HOSTS>
      <HOST id="A" address="129.127.8.34" />
      <HOST id="B" address ="129.127.8.35" />
   </HOSTS>
   <DEPLOYMENTS>
      <DEPLOYMENT name="PrimaryServer" bundle="Server" target="A" />
      <DEPLOYMENT name="CachingServer" bundle="Cache" target="B" />
   </DEPLOYMENTS>
   <CONNECTIONS>
      <CONNECTION>
         <SOURCE       deployment="PrimaryServer"
                       channel="DownstreamCache" />
         <DESTINATION  deployment="CachingServer"
                       channel="UpstreamServer" />
      </CONNECTION>
   </CONNECTIONS>
</DDD>
```

**Fig 4**: A Deployment Description Document

The first phase of application deployment involves installation of component bundles on appropriate hosts. The deployment engine reads a DDD and retrieves the specified bundles from their given locations (which may be within a local file-based component catalogue or elsewhere on the network). It then configures an installer bundle for each host by generating an appropriate to-do list. These installers are fired as illustrated in **Fig 1** on the participating thin servers throughout the network. The action of each executing installer bundle on arrival is to extract its payload component bundles and add them to the local store. It then sends a task report back to the deployment engine listing the resulting store keys of the installed bundles, and terminates.

The second phase of application deployment involves starting execution of the previously installed 'dormant' component bundles on the appropriate thin servers. The deployment engine configures a runner bundle for each thin server, by generating an appropriate to-do list, and fires it on that thin server. The action of each executing runner bundle on arrival is to extract the relevant bundle(s) from the local store, and to fire these in turn. **Fig 2** shows an example runner bundle, containing the store key (labelled as *StoreGuid*) of the bundle to be extracted and fired. As with the installation process, the runner sends a task report back to the deployment engine, in this case listing the connectors of the enclosing machine for each fired bundle. This will enable the deployment engine to communicate with the newly running bundles in the final wiring phase. The process is illustrated in **Fig 5**, in which a runner is fired on thin server *A*. The runner retrieves the *Server* bundle from *A*'s store and fires it in a new machine, and returns a task report to the deployment engine.

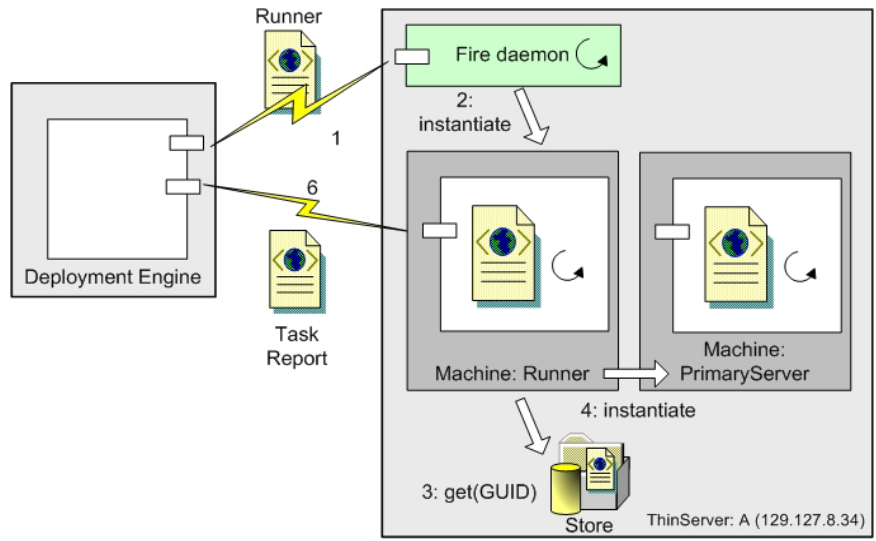

**Fig 5**: The running process

The final phase of application deployment involves connecting (termed *wiring*) the named channels on the running bundles to assemble the global application topology. The deployment engine configures a wirer bundle for each connection, by generating an appropriate to-do list. The wiring process will begin on one of the thin servers selected arbitrarily. Each wirer created is configured with a to-do list describing:

1. The connector for each machine—this contains the IP address of the machine and the machine and resource ports.
2. The name used by the executing bundle to reference the channel in both machines (this may be different for each machine).

**Fig 6** shows an example wirer bundle for thin server *A*, which is (arbitrarily) chosen as the initiating thin server for the connection between the named channels *DownstreamCache* and *UpstreamServer*. Note that this bundle cannot be generated by the deployment engine until it has received the task reports from the runner bundles, since those contain the necessary port numbers.

The executing wirer bundle is able to communicate with the relevant machine's connection manager via its machine channel. In the example, the wirer executing on thin server *A* requests that the machine containing the running *Server* bundle create a new named channel with the name *DownstreamCache*. In response to this request, the machine's connection manager also starts a thread listening for incoming TCP/IP socket connections, and returns the port number to the wirer.

The wiring must now be completed by having the machine containing the running *Cache* bundle on thin server *B* connect to *A* on that port. To achieve this, the wirer on *A* configures another wirer bundle (its 'offspring'), which is fired on *B*. The purpose of the offspring wirer is to connect the named channel on *B* to the waiting channel on *A*. When the offspring wirer arrives at *B*, it communicates with the connection manager of the appropriate machine and instructs it to create a new named channel *Up*- *streamServer* and connect it to the *DownstreamCache* channel by communicating with the listening port on *A*, thus establishing the connection. This process is illustrated in **Fig 7.**

```xml
<BUNDLE>
   <AUTHENTICATION entity="1973073447eb91509"
                   signature=" CS68m..+SLGU" />
   <CODE entry="uk.ac.stand.cingal.Wirer" type="java">
      <CLASS name="uk.ac.stand.cingal.Wirer">
         sdjskF2YS9GFGSDnL09fdsa…
      </CLASS>
   </CODE>
   <DATA>
      <DATUM id="ToDoList">
         <TODOLIST>
            <TASK guid="urn:cingal:322xf344" type="WIRE">
               <DATUM id="PrimaryConnector">
                  <CONNECTOR  host="129.127.8.34"
                              machinePort="30112"
                              resourcePort="29000" /></DATUM>
               <DATUM id="SecondaryConnector">
                  <CONNECTOR  host="129.127.8.35"
                              machinePort="47121"
                              resourcePort="26083" /></DATUM>
               <DATUM id="PrimaryNamedChannel">
                  DownstreamCache</DATUM>
               <DATUM id="SecondaryNamedChannel">
                  UpstreamServer</DATUM>
            </TASK>
         </TODOLIST>
      </DATUM>
   </DATA>
</BUNDLE>
```

**Fig 6**: A wirer bundle

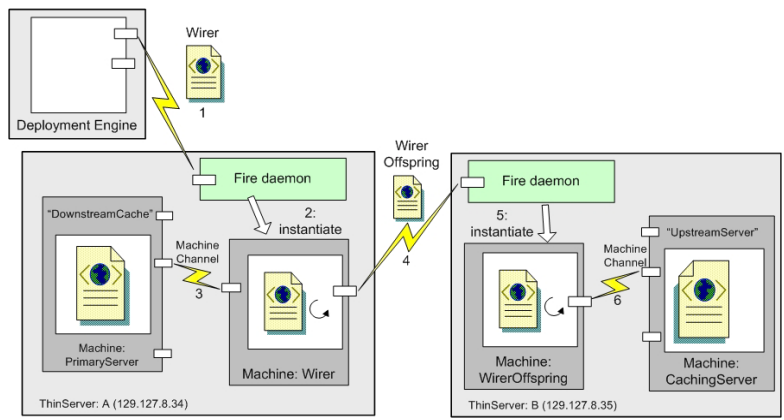

**Fig 7**: The wiring process

Once all the wirers have completed their (possibly parallel) computation, the wiring process is complete and the named channels are connected as shown in **Fig 8**. This completes the deployment of the distributed application.

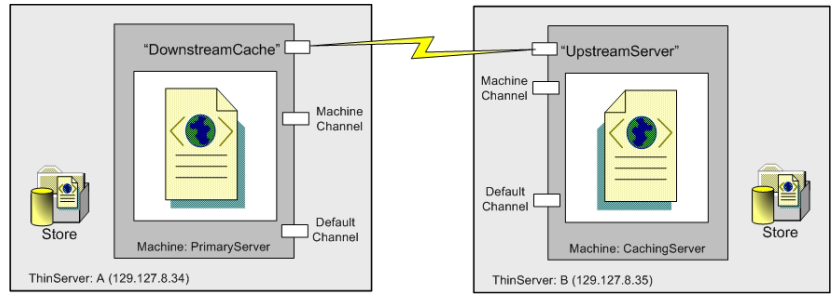

**Fig 8**: Result of wiring process

## 4 Related Work

The OSGi Service Platform [6] has perhaps the most in common with this work; it addresses similar issues of remote installation and management of software components, and (independently) adopts similar terminology for *bundles* and *wiring*. The most significant difference is the lack of high-level declarative architectural descriptions. This arises from it being targeted primarily at software deployment onto smart and embedded devices, whereas Cingal is aimed more generally at deployment and evolution of distributed applications on the basis of explicit architectural descriptions. As mentioned in Section 6, we are currently working on generating these descriptions automatically from high-level goals specified as constraints, to allow automatic reconfiguration and deployment in response to observed run-time problems such as host or network failure. Another difference is in the wiring model: a given OSGi bundle can be a producer and/or a consumer, and all its associated wires are conceptually equivalent. Cingal allows any number of symbolically named ports to be associated with a bundle, and the programmer may treat these differently. However, the two schemes have equivalent modelling power. Finally, Cingal is more flexible with regards to initial provisioning: its ubiquitous *fire* service allows bundles to be pushed to a new node from a remote management agent without any intervention required locally on the node. Initial provisioning in OSGi involves pull from a new node, which must be initialised somehow with an address from which to pull the code. The address may be provided by various means such as direct user intervention, factory installation, reading from a smartcard, etc.

A number of languages have been developed to describe software architectures, including [7-9]. Typical of these is Acme [1], which is intended to fulfil three roles: to provide an architectural interchange format for design tools, to provide a foundation for the design of new tools and to support architectural modelling. The Acme language supports the description of components joined via connectors, which pro-vide a variety of communication styles. Components and connectors may be annotated with properties that specify attributes such as source files and degrees of concurrency, etc. Acme also supports a logical formalism based on relations and constraints, which permits computational or run-time behaviour to be associated with the description of architectures. Acme does not support the deployment of systems from the architectural descriptions, nor does it encompass physical computation resources.

The ArchWare ADL [10] is based on higher-order π-calculus, and is aimed at specifying active architectures, in which the architectural description of an application evolves in lock-step with the application itself. The language supports a reversible *compose* operator that allows components to be assembled from other components, and later decomposed and recomposed to permit evolution. Decomposition operates at a fine-grain, and it is possible to decompose a component into constituent parts without losing encapsulated state. This is achieved using *hyper-code* [11], which provides a reified form for both code and data. In comparison, the ArchWare ADL focuses on software architecture and does not address physical deployment.

The idea of installing components using agents has its roots in a number of places. Java Servlets were initially developed within Sun Labs with the express purpose of freeing the restrictions of a fixed repertoire service, allowing the client to modify the behaviour of the server. However, as they moved into a commercial product domain, this flexibility was removed as it was deemed to compromise the security of the underlying traditional operating system [12]. The Infospheres project from Caltech [13] has some overlap with the system described here. They propose a system of distributed Java processes (dapplets), which can be connected using asynchronous message passing.

The Tacoma system [14] uses agent technology to install software on remote machines and like our system uses digital signatures to verify the authenticity of agents. Tacoma introduces the notions of a *briefcase* to carry agent payloads, which may include components, and a *cabinet*, which is a persistent site bound briefcase, corresponding closely to our *store*. However, the Tacoma system appears to be aimed at installing non-distributed applications on remote nodes and does not include the notion of distributed components, communication channels nor high-level architectural descriptions.

## 5 Conclusions

At the start of this paper six claims were made about our deployment architecture[3]. In conclusion, these claims are critically re-examined.

**Claim 1**: The Cingal infrastructure permits bundles to be deployed in arbitrary geographic locations from conventional machines. Bundles may perform arbitrary computation and offer arbitrary network services.

---

[3] The current implementation may be downloaded from
   *http://www-systems.dcs.st-and.ac.uk/cingal/downloads/*.

**Claim 2**: The runtime infrastructure provided by Cingal thin servers abstracts over host-specific differences yielding a homogeneous run-time environment for deployed components.

**Claim 3**: The store and binder provided by thin servers support content-addressed storage, which permits code and data to be stored with no possibility of ambiguous retrieval. The binder permits objects to be symbolically named to facilitate the retrieval of components whose content keys are not known. The binder also provides an evolution point supporting update of component mappings.

**Claim 4**: Deployment Description Documents support the specification of distributed architectures. The deployment engine technology combined with the thin server infrastructure permits these distributed deployments to be realised into running instances of component based architectures. The process of deployment from specification through to having a connected collection of running components on distributed hosts is totally automated.

**Claim 5**: A number of novel evolution mechanisms are provided by the architecture. Firstly, the architecture supports the ability to remotely update components. Secondly flexible binding between components is made possible thorough the binder and store interfaces. Most importantly, distributed architectures may be re-arranged by unbinding and reconnecting named channels within machines running on thin servers.

**Claim 6**: The two security mechanisms provided by Cingal prevent unauthorised entities from firing bundles on hosts on which they do not have privilege. The ownership model which makes uses of standard cryptographic certificate techniques is well suited to distributed deployment. Tools (not described here) that operate in a similar manner to the deployment tool are provided for managing entity privileges and updating collections of machines. The capability protection system provided within Cingal thin servers prevents bundles being used for malicious or unintentional abuse of the thin server infrastructure.

## 6 Future Work

In the future we propose to expand the system in two primary ways. Firstly we would like to make the specification of distributed components more declarative. To this end we are currently investigating the use of constraint based specification languages. It is our intention to construct higher level specifications and a set of tools to support them and compile these specifications down to DDD documents. Secondly, we are investigating how evolution can be specified at the DDD level. Since we use DDDs to specify deployments, it seems natural to have high-level descriptions of evolution and automatically generate bundles to enact the necessary changes. Another interesting line of investigation is the use of a higher level specification of architectural intent from which DDDs may be generated. We are currently investigating the use of constraint based specification languages for this purpose [5]. We believe that this approach may be combined with the infrastructure described in this paper to yield systems that are capable of autonomic evolution in the face of perturbations such as host and link failure, temporary bandwidth problems, etc. We postulate that it will be feasible to implement an autonomic manager that will automatically evolve the deployed application to maintain the constraints while it is in operation.

# 7 Acknowledgements

This work is supported by EPSRC grants GR/M78403 "Supporting Internet Computation in Arbitrary Geographical Locations", GR/R51872 "Reflective Application Framework for Distributed Architectures" and by EC Framework V IST-2001-32360 "ArchWare: Architecting Evolvable Software".

The authors thank Richard Connor for his invaluable input to the Cingal project, on which he was a principal investigator, and Ron Morrison who read early drafts of this paper.